\documentclass{elsart}
\usepackage{amssymb}
\usepackage{amsmath}
\usepackage{rotating}

\input{tcilatex}

\begin{document}

\begin{frontmatter}

\title{ Entropy and Uncertainty Analysis in Financial Markets}

\author{*Andreia Dionisio, **Rui Menezes and **Diana A. Mendes}

\address{*University of Evora, Center of Business Studies, CEFAGE-UE, Largo Colegiais, 2, 7000 Evora, Portugal, E-mail: andreia@uevora.pt;
 **ISCTE, Department of Quantitative Methods, Av. Forcas Armadas, 
 1649-Lisboa, Portugal, E-mail: rui.menezes@iscte.pt, diana.mendes@iscte.pt}

\begin{abstract}
The investor is interested in the expected return and he is also concerned about the risk and the uncertainty assumed by the investment. One of the most popular concepts used to measure the risk and the uncertainty is the variance and/or the standard-deviation. In this paper we explore the following issues: Is the standard-deviation a good measure of risk and uncertainty? What are the potentialities of the entropy in this context? Can entropy present some advantages as a measure of uncertainty and simultaneously verify some basic assumptions of the portfolio management theory, namely the effect of diversification?
\end{abstract}


\end{frontmatter}

\section*{Introduction}

This paper examines the adequacy of entropy as a measure of uncertainty in
portfolio management in finance and its behaviour is compared with the most
popular risk measure used in finance: the variance.

It is quite common to relate the variance or the standard-deviation and the
VaR (Value-at-Risk) as the main measures of risk and uncertainty in finance.
However, some authors [see \emph{e.g.} Soofi (1997)] point out that these
measures may fail in some specific situations as measures of uncertainty,
since they require that the underlying probability distribution is symmetric
and neglect the possibility of extreme events such as, for example, the
existence of fat-tails.

The main goal of this paper is to assess the capability of entropy to
measure the uncertainty in portfolio management. We describe the theoretical
background of entropy and its mathematical properties, we present a
comparative analysis between the entropy and the variance/standard-deviation
as measures of uncertainty in the stock market and we discuss the empirical
results of a comparative analysis between the CAPM and some measures of
information theory (namely entropy, conditional entropy and mutual
information).

The results obtained point to the conclusion that the entropy observes the
effect of diversification and is a more general measure of uncertainty than
the variance, since it uses more information about the probability
distribution. The mutual information and the conditional entropy show a good
performance when compared with the systematic risk and the specific risk
estimated through the linear Market Model.

\section{Theoretical Background}

According to Shannon (1948) entropy satisfies the main properties of a good
measure of uncertainty. Let $p_{1},...,p_{n}$ be the probabilities of
occurrence of a set of events. The entropy for discrete distributions is
given by [Shannon (1948)] $H\left( X\right) =-\sum\nolimits_{i}p_{i}\log
p_{i}.$ For continuous distributions, where $p_{X}(x)$ is the density
function of the random variable $X$, the entropy is given by 
\begin{equation}
H\left( X\right) =-\int p_{X}(x)\log p_{X}(x)dx.  \label{continuous}
\end{equation}

Note that $H\left( X\right) $ and $H\left( Y\right) $ are the entropies of
the random variables $X\in \overrightarrow{X}$ and $Y\in \overrightarrow{Y}$%
, $H\left( X,Y\right) $ is the joint entropy, and $H\left( Y|X\right) $ and $%
H\left( X|Y\right) $ are the conditional entropies.

The properties of entropy for discrete and continuous distributions are
mainly alike. In particular we have [Shannon (1948); Kraskov\emph{\ et al.}
(2004)]: $\left( i\right) $ if $X$ is limited to a certain volume $v$ in its
space, then $H(X)$ is a maximum and is equal to $\log v$ when $p_{X}(x)$ is
constant, $1/v$, in the volume; $\left( ii\right) $ for any two variables $X$
and $Y$, we have $H\left( X,Y\right) \leq H\left( X\right) +H\left( Y\right)
,$ where the equality holds if (and only if) $X$ and $Y$ are statistically
independent, \emph{i.e.} $p_{X,Y}(x,y)=p_{X}(x)p_{Y}(y);$ $\left( iii\right) 
$\ the joint entropy is given by $H\left( X,Y\right) =H\left( X\right)
+H\left( Y|X\right) =H\left( Y\right) +H\left( X|Y\right) ,$since $H\left(
X\right) +H\left( Y\right) \geq H\left( X,Y\right) ,$ then $H\left( Y\right)
\geq H\left( Y|X\right) $ and $H\left( X\right) \geq H\left( X|Y\right) .$

The entropy of a normal distribution can be parametrically estimated by $%
NH\left( X\right) =\log \left( \sqrt{2\pi e}\sigma \right) ,$ where $\sigma $
is the standard-deviation, and $\pi $ and $e$ are defined as usually. Notice
that the assumption that the data follow a normal distribution is very
common in portfolio management and regression analysis.

The mutual information is a measure of association between variables and can
be defined as follows [Shannon (1948)]:%
\begin{equation}
I\left( X,Y\right) =H\left( Y\right) -H\left( Y|X\right) =\int \int
p_{X,Y}\left( x,y\right) \log \frac{p_{X,Y}\left( x,y\right) }{p_{X}\left(
x\right) p_{Y}\left( y\right) }dxdy.  \label{MI}
\end{equation}%
The mutual information is a nonnegative measure [Kullback (1968)], being
equal to zero if and only if $X$ and $Y$ are statistically independent. In
this way, the mutual information between two random variables $X$ and $Y$
can be seen as a measure of dependence between these variables, or even
better, it can be regarded as a measure of the statistical correlation
between $X$ and $Y$. However, we can not say that $X$ is causing $Y$ or
vice-versa.

We also need to define a measure that can be directly comparable with the
linear correlation coefficient. In equation $\left( \ref{MI}\right) $, we
have $0\leq I\left( X,Y\right) \leq +\infty $, which hampers eventual
comparisons between different samples. Some authors, namely Granger and Lin
(1994), Darbellay (1998) and Soofi (1997) have used a standardized measure
for the mutual information, referred to as the global correlation
coefficient, defined by $\lambda \left( X,Y\right) =\sqrt{1-e^{-2I\left(
X,Y\right) }}.$ This measure varies between $0$ and $1,$ being thus directly
comparable with the linear correlation coefficient.

The function $\lambda \left( X,Y\right) $ captures the overall linear and
nonlinear dependence between $X$ and $Y$.$\ $This measure can be used as a
measure of predictability based on an empirical probability distribution,
although it is not derived from any particular model of predictability. In
this particular case, the above mentioned properties assume the following
form: $\left( i\right) \ \lambda \left( X,Y\right) =0$, if and only if $X$
contains no information on $Y;$ $\left( ii\right) \ \lambda \left(
X,Y\right) =1$, if there is a perfect relationship between the vectors $X$
and $Y$. This is the farthest case of determinism; $\left( iii\right) $\
when modelling the input-output pair $\left( X,Y\right) $ by any model with
input $X$ and output $U=f\left( X\right) $, where $f$ is a function of $X$,
the predictability of $Y$ by $U$ cannot exceed the predictability of $Y$ by $%
X,$ \emph{i.e.}, $\lambda \left( X,Y\right) \geq \lambda \left( U,Y\right) .$

One of the difficulties to estimate the mutual information on the basis of
empirical data lies on the fact that the underlying \emph{p.d.f. }is
unknown. To overcome this problem, there are essentially three different
methods to estimate the mutual information: \ histogram-based estimators,\
kernel-based estimators and\ parametric methods.\footnote{%
The histogram-based estimators can be divided in two groups: equidistant
cells and equiprobable cells, \emph{i.e.} marginal equiquantisation. The
second approach presents some advantages, since it allows for a better
adequacy to the data and maximizes mutual information [Darbellay (1998)].}
We will use the marginal equiquantization histogram-based estimation process
proposed by Darbellay (1998), in order to minimize the bias that may occur.

Finally, we should note that the introduction of entropy as a measure of
uncertainty in finance goes back to Philippatos and Wilson (1972), who
present a comparative analysis between the behaviour of the
standard-deviation and the entropy in portfolio management. They conclude
that entropy is more general and has some advantages relatively to the
standard-deviation, such as Dionisio \emph{et al.} (2006).

\section{Entropy and variance: a comparative analysis}

Usually, the variance is the central measure in the risk and uncertainty
analysis in financial markets. However, the entropy can be used as an
alternative measure of dispersion, and some authors consider that the
variance should be interpreted as a measure of uncertainty with some
precaution [see, \emph{e.g.} Maasoumi (1993) and Soofi (1997)].

Ebrahimi, Maasoumi and Soofi (1999) examined the role of \ the variance and
entropy in ordering distributions and random prospects, and concluded that
there is no general relationship between these measures in terms of ordering
distributions. They found that, under certain conditions, the ordering of
the variance and entropy is similar for transformations of continuous
variables, and show that the entropy depends on many more parameters of a
distribution than the variance. Indeed, a Legendre series expansion shows
that the entropy is related to higher-order moments of a distribution and
thus, unlike the variance, could offer a better characterization of $%
p_{X}\left( x\right) $ since it uses more information about the probability
distribution than the variance [see Ebrahimi\emph{\ et al.} (1999)].

Maasoumi and Racine (2002) argue that when the empirical probability
distribution is not perfectly known, the entropy constitutes an alternative
measure for assessing the uncertainty, predictability and also
goodness-of-fit. Likewise, McCauley (2003) defends that the entropy
represents the disorder and uncertainty of a stock market or a particular
stock, since the entropy has the ability to capture the complexity of
systems without requiring rigid assumptions that may bias the results
obtained.

Our empirical analysis is based on daily closing prices of 23 stocks rated
in the Portuguese stock market (\emph{Euronext Lisbon}), covering the period
from June, 28, 1995 to December, 30, 2002, which corresponds to 1858
observations \emph{per} stock in order to compute the rates of return. The
stock index PSI 20 is used as the market benchmark, or proxy, since it is
the index that better represents the \emph{Euronext Lisbon}. A preliminary
statistical analysis of the rates of return reveals that the null that the
empirical distributions are Gaussian should be rejected since they show high
levels of kurtosis and skewness.

Firstly, we perform a comparative analysis between the entropy and the
logarithm of the standard-deviation, $\ln \left( \sigma \right) $, for each
stock in our data set and for the stock index PSI 20. The entropy was
computed using expression (\ref{continuous}) measured in $nats$. The $\ln
\left( \sigma \right) $\ was used instead of\ $\sigma $\ in order to provide
a correct comparison between these measures. The results are shown in Figure %
\ref{Fig1}.

\FRAME{fhFU}{3.1799in}{2.5417in}{0pt}{\Qcb{Entropy \emph{versus} $\ln \left( 
\protect\sigma \right) .$}}{\Qlb{Fig1}}{dionisiofig1.eps}{\special{language
"Scientific Word";type "GRAPHIC";maintain-aspect-ratio TRUE;display
"USEDEF";valid_file "F";width 3.1799in;height 2.5417in;depth
0pt;original-width 6.7914in;original-height 5.4232in;cropleft "0";croptop
"1";cropright "1";cropbottom "0";filename '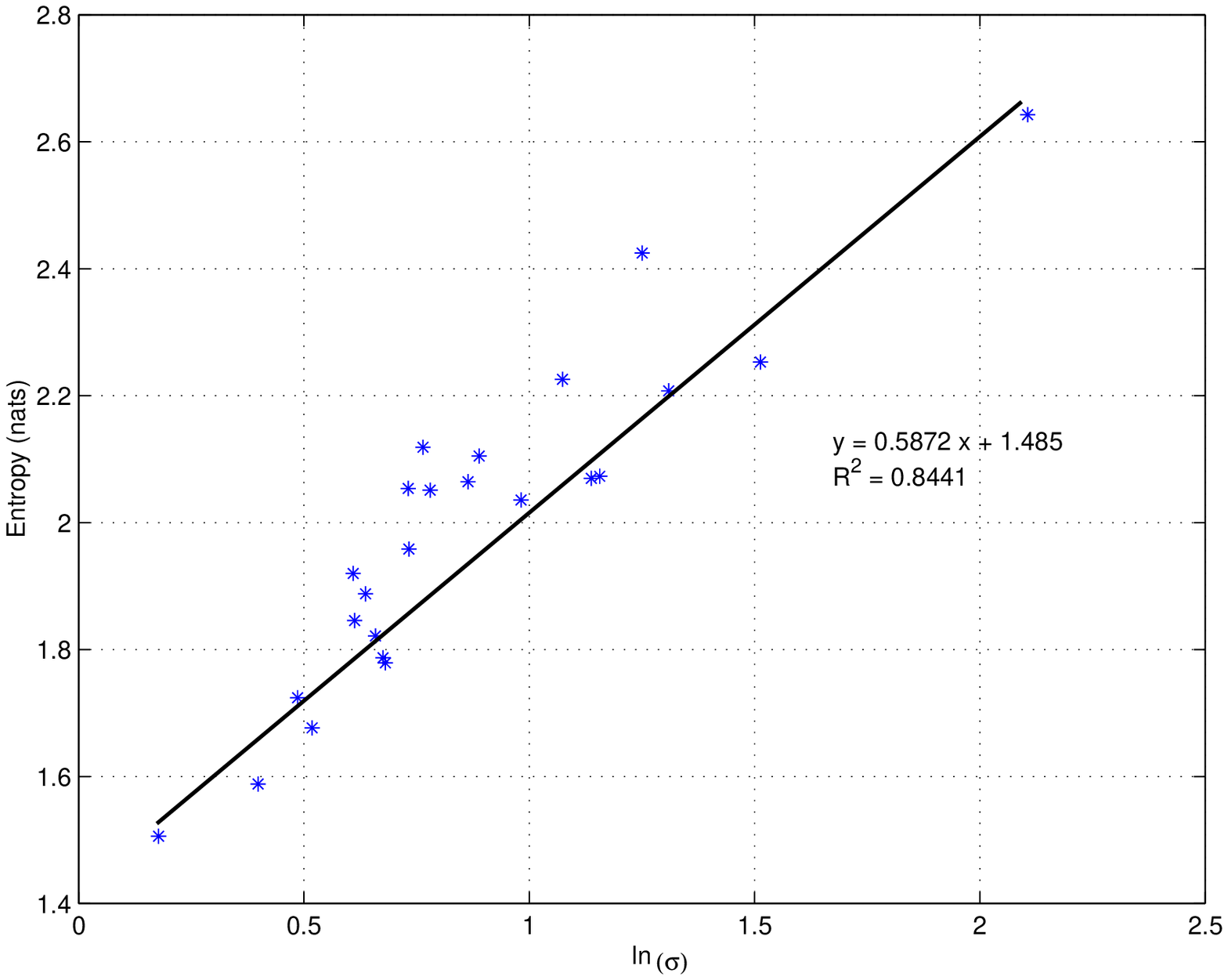';file-properties
"XNPEU";}}

As may be seen, there is a strong positive relationship between the entropy
and the $\ln \left( \sigma \right) $. There are, however, some larger
deviations and the null under the Jarque-Bera test is rejected for all
stocks and also for the stock index PSI 20. In Figure \ref{Fig1} the larger
deviations from the regression line correspond to stocks that also exhibit
higher levels of skewness and kurtosis.

We also compared the results between the empirical entropy and the normal
entropy for each stock. The results reveal that the normal entropy takes
always higher values than the empirical entropy, indicating that the
uncertainty of these stocks and index is smaller than that we would observe
if they were normally distributed. Thus, we can admit evidence of some
predictability of the rate of returns, or at least that it is higher than
the one assumed by the financial theory. Again, the main differences between
the normal entropy and the empirical entropy are found in the stocks that
exhibit the highest levels of kurtosis, skewness, autocorrelation and
heteroskedasticity. Therefore, the empirical entropy appears to be sensitive
to higher-order moments of the distribution, supplying thus more information
about the stock and its probability distribution.\footnote{%
We also performed an analysis of the diversification effect for randomly
selected portfolios [see Dionisio \emph{et al.} (2006)]. This study revealed
that both the entropy and the standard-deviation tend to decrease when more
assets are included in the portfolio, pointing to the conclusion that
entropy is sensitive to the effect of diversification.}

\section{Analysis of dependency between the stocks and the PSI 20 index}

In the context of portfolio management researchers tend to pay more
attention to the systematic risk than to the specific risk, because the
latter can be minimized (and in the limit can be zero) by an effective
diversification process. Usually, the systematic risk in financial theory is
measured by the \emph{Beta} of the CAPM. In this context, it is assumed that
the rate of returns of an asset $i$ is equal to the sum of the risk free
rate of return $\left( R_{f}\right) $ and the compensation for the risk $%
\left\{ \left[ E\left( R_{m}\right) -\left( R_{f}\right) \right] \beta
_{i}\right\} $. In this second term, the coefficient $\beta $ plays an
important role, since it measures the sensibility of the rate of returns of
the asset (or portfolio) to the risk premium, \emph{i.e.} the systematic
risk. Recall that the variance of an asset or portfolio $\left( \sigma
_{i}^{2}\right) $ can be decomposed in two components: the sum of squares of
the regression and the residual sum of squares, that is%
\begin{equation}
\sigma _{i}^{2}=\beta _{i}^{2}\sigma _{m}^{2}+\sigma _{\epsilon i}^{2},
\label{var-beta}
\end{equation}%
where $\sigma _{m}^{2}$ is the variance of the independent variable, in this
case the variance of the market benchmark, and $\sigma _{\epsilon i}^{2}$ is
the residual term of the variance that can be minimized through a
diversification process. In order to estimate the \emph{Beta} of the CAPM we
use the Market Model that can be given by 
\begin{equation}
R_{it}=R_{_{ft}}+\left[ R_{mt}-R_{ft}\right] \beta _{i}+\varepsilon _{it}.
\end{equation}%
The Market Model is usually estimated by OLS. However, statistical tests and
the empirical evidence show that the residuals are not white noise and thus
OLS may not be appropriate.

The main goal of this section is to evaluate the level of dependence between
each stock and the stock index. The measures of information theory described
before, namely the entropy, the conditional entropy and the mutual
information are used to evaluate the (in)dependence between the stocks and
the stock index PSI 20. If the residuals are white noise, then the global
correlation coefficient $\left( \lambda \right) $ will be similar to the
linear correlation coefficient $(R)$ and the \emph{Beta} is a good measure
of the systematic risk. However, in case of nonlinearities and
irregularities in the behaviour of the residuals, the simple linear
regression model is not able to capture the existence of a global
relationship between the stocks and the stock index PSI 20. In this case,
the mutual information and the global correlation coefficient can be
potential sources of information for the investor.

One may distinguish the "global" uncertainty from the "residual" uncertainty
based on the properties of the entropy. The entropy of a stock (or any other
variable) can be decomposed as follows%
\begin{equation}
H\left( X\right) =I\left( X,PSI\right) +H\left( X|PSI\right) ,
\label{ent-bet-eq}
\end{equation}%
where $X$ represents the stock and $PSI$ denotes the stock index. The
expression $\left( \ref{ent-bet-eq}\right) $ can be roughly compared with
the expression $\left( \ref{var-beta}\right) $, where the first term refers
to the level of association or dependence between the asset (stock or
portfolio) and the proxy $PSI$, and the second term is the variation of that
asset (or stock, or portfolio) that is independent from the proxy used.

The \emph{Beta} was estimated for each stock as well as the linear
correlation coefficient resulting from the Market Model. In addition, we
also calculate the systematic risk, $\beta _{i}^{2}\sigma _{m}^{2},$ and the
specific risk, $\sigma _{\epsilon i}^{2}.$\ In order to provide some
comparisons between the two approaches, we calculated the mutual information
between each stock and the stock index PSI 20, $\left[ I\left( X,PSI\right) %
\right] ,$ the conditional entropy $\left[ H\left( X|PSI\right) \right] ,$
and the global correlation coefficient, $\lambda .$

As we can see in Figure \ref{Figure2}, there is a positive relationship
between the systematic risk and the mutual information, and between the
specific risk and the conditional entropy, although these measures are not
directly comparable.

\FRAME{fhFU}{4.126in}{2.9395in}{0pt}{\Qcb{Comparative analysis between the
systematic risk, $\protect\beta _{i}^{2}\protect\sigma _{m}^{2}$ (continuous
line)$,$ and the mutual information $I$ and between the specific risk, $%
\protect\sigma _{\protect\varepsilon i}^{2}$ (continuous line)$,$ and the
conditional entropy, $H\left( X|PSI\right) .$}}{\Qlb{Figure2}}{%
dionisiofig2.eps}{\special{language "Scientific Word";type
"GRAPHIC";maintain-aspect-ratio TRUE;display "USEDEF";valid_file "F";width
4.126in;height 2.9395in;depth 0pt;original-width 7.3907in;original-height
5.2581in;cropleft "0";croptop "1";cropright "1";cropbottom "0";filename
'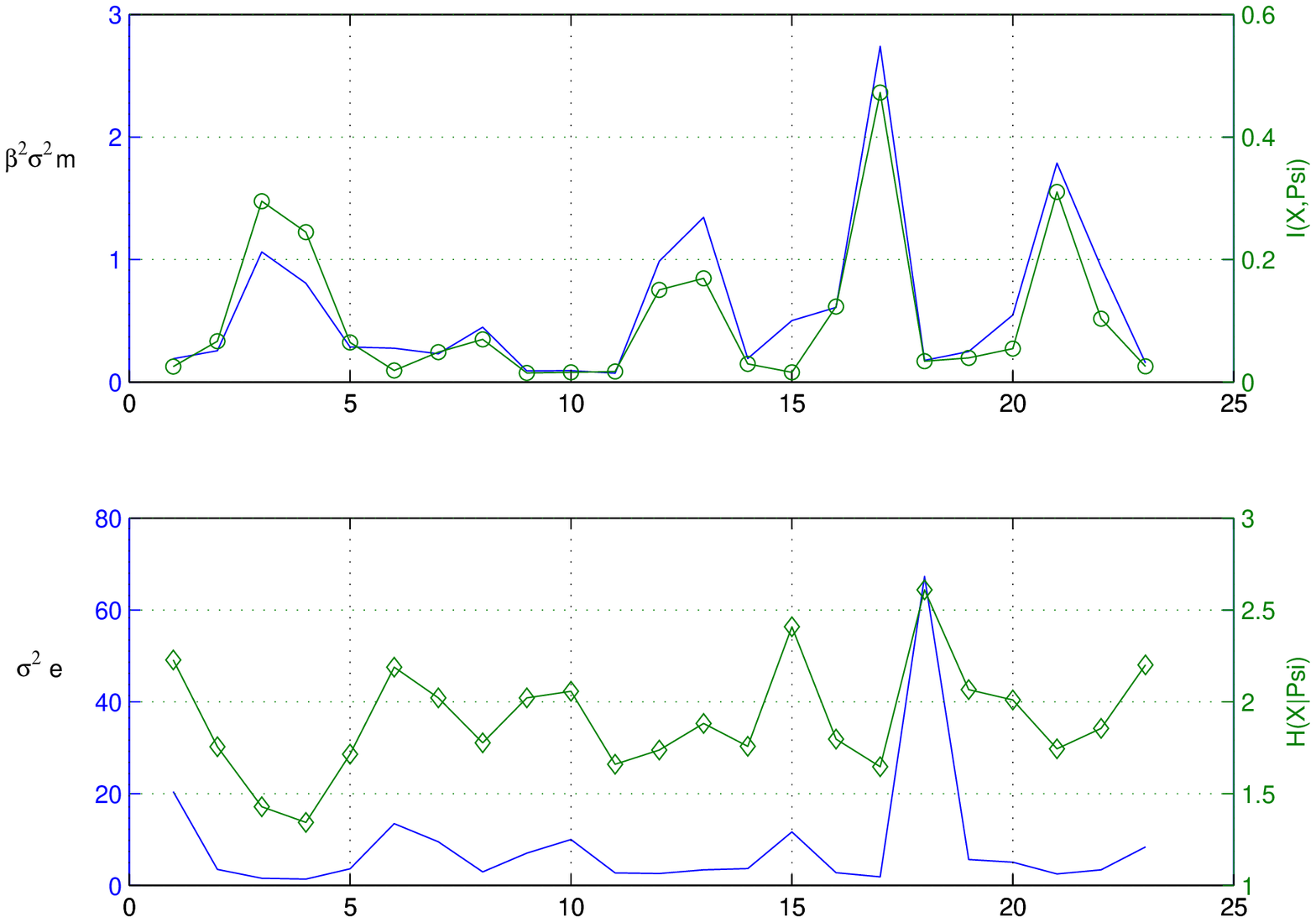';file-properties "XNPEU";}}

In spite of the evidence of a positive and strong relationship between the
variance and the measures of the theory of the information, we performed a
comparison between measures that can be directly compared. To this end, we
used the global correlation coefficient or \emph{lambda} $\left( \lambda
\right) $ and the global correlation coefficient assuming that the
distribution is normal or \emph{lambda normal} $\left( \lambda _{n}\right) $.%
\footnote{%
The lambda normal was computed using equation the concept of normal mutual
information given by: $IMN\left( X,Y\right) =-\frac{1}{2}\log \left(
1-R^{2}\left( X,Y\right) \right) $, where $R^{2}$ is the coefficient of
determination.} We found that there are stocks whose relationship with the
stock index PSI 20 exhibit strong discrepancies when analyzed from a global
or purely linear perspective. In order to find the possible causes of such
differences, several tests were accomplished for the residuals produced by
the estimation of the linear Market Model, namely the Ljung-Box test, the
Jarque-Bera test, the Engle test and the stability tests \emph{CUSUM} and 
\emph{CUSUM-Q}. The results of these tests indicate that there are precisely
the residuals resulting from the application of the linear Market Model to
stocks that appear to exhibit higher levels of nonlinearity that present
more evidence of autocorrelation, non-normality, heteroskedasticity and
nonstability, and this seems to be an indicator that linear analyses are
possibly not enough to evaluate risk and uncertainty.

\section{Conclusions}

The results presented in this paper point to the conclusion that the entropy
observes the effect of diversification and is a more general measure of
uncertainty than the variance, since it uses more information about the
probability distribution. The mutual information and the conditional entropy
show a good performance when compared with the systematic risk and the
specific risk estimated through the linear Market Model. Nevertheless, the
use of the concept of entropy in risk analysis and portfolio selection needs
some care, because it does not take into account the actual values of the
variables and this may eventually compromise its inclusion in the context of
an utility function.\medskip

\bigskip \textbf{Acknowledgement}

Financial support from Fundacao da Ciencia e Tecnologia, Lisbon, is
gratefully acknowledged by the authors, under the contract
PTDC/GES/70529/2006.


\begin{thebibliography}{99}
\bibitem{} {\small Darbellay, G. Predictability: an Information-Theoretic
Perspective. In: Signal Analysis and Prediction, A. Proch\'{a}zka, J. Uhl%
\'{\i}r, P.J.W. Rayner and N.G. Kingsbury, Birkhauser eds., Boston, \
(1998), 249-262.}

\bibitem{} {\small Dionisio, A., Menezes, R. and Mendes, D.A. The European
Physical Journal B, (2006), 50, 161-164.}

\bibitem{} {\small Ebrahimi, N., Maasoumi, E., Soofi, E. Journal of
Econometrics, (1999), 90, 2, 317-336.}

\bibitem{} {\small Granger, C., Lin, J. Journal of Time Series Analysis,\
(1994), 15, 4, 371-384.}

\bibitem{} {\small Kraskov, A., St}\"{o}{\small gbauer H., Andrzejak, R.,
Grassberger, P., Hierarchical Clustering Based on Mutual Information,
(2004), preprint http://www.arxiv:q-bio.QM/0311039.}

\bibitem{} {\small Kullback, S. Information Theory and Statistics. Dover,
New York. 1968.}

\bibitem{} {\small Maasoumi, E. Econometric Reviews, (1993), 12, 2, 137-181.}

\bibitem{} {\small Maasoumi, E., Racine, J. Journal of Econometrics, (2002),
107, 291-312.}

\bibitem{} {\small McCauley, J. Physica A, (2003), 329, 199-212.}

\bibitem{} {\small Philippatos, G., Wilson,\ C. Applied Economics, (1972),
4, 209-220.}

\bibitem{} {\small Shannon, C. E. Bell Systems Tech., (1948), 27: 379-423,
623-656.}

\bibitem{} {\small Soofi, E.. Information Theoretic Regression Methods. In:
Advances in Econometrics - Applying Maximum Entropy to Econometric Problems,
Fomby, T. and R. Carter Hill eds. Vol. 12. Jai Press Inc., Lo}n{\small don.
1997.}
\end{thebibliography}
\end{document}